# 3D sensors for the HL-LHC

D. Vázquez Furelos[a,1], M. Carulla[c], E. Cavallaro[a], F. Förster[a], S. Grinstein[a,b],
J. Lange[a], I. López Paz[a], M. Manna[a], G. Pellegrini[c], D. Quirion[c] and S. Terzo[a]

[a] *Institut de Física d'Altes Energies (IFAE), The Barcelona Institute of Science and Technology (BIST),*
   *Campus UAB, Edifici CN, E-08193 Bellaterra (Barcelona), Spain*
[b] *Institució Catalana de Recerca i Estudis Avançats (ICREA),*
   *Passeig Lluís Companys 23, E-08010 Barcelona, Spain*
[c] *Centro Nacional de Microelectrónica (CNM-IMB-CSIC),*
   *Campus UAB, Carrer dels Til·lers, E-08193 Bellaterra (Barcelona), Spain*

*E-mail*: david.furelos@cern.ch

ABSTRACT: In order to increase its discovery potential, the Large Hadron Collider (LHC) accelerator will be upgraded in the next decade. The high luminosity LHC (HL-LHC) period demands new sensor technologies to cope with increasing radiation fluences and particle rates. The ATLAS experiment will replace the entire inner tracking detector with a completely new silicon-only system. 3D pixel sensors are promising candidates for the innermost layers of the Pixel detector due to their excellent radiation hardness at low operation voltages and low power dissipation at moderate temperatures. Recent developments of 3D sensors for the HL-LHC are presented.

KEYWORDS: HL-LHC; ATLAS upgrade; High Luminosity, phase 2 LHC.

---

[1] Corresponding author

# Contents

## 1. 3D sensors in ATLAS

The ATLAS experiment relies on the performance of silicon pixel detectors to identify and determine the paths of particles that are produced in the LHC proton-proton collisions. The rate of the particles and the radiation levels produced put heavy constraints on the sensor and front-end electronics that are part of the pixel system. An upgrade of the LHC accelerator into the High Luminosity LHC (HL-LHC) is planned for 2024, which will deliver five to ten times the nominal instantaneous luminosity of the LHC and an integrated luminosity up to 3000 fb$^{-1}$. To maintain the detector performance new pixel technologies have to be developed. Recently, 3D sensors have emerged as a viable solution for the innermost layers of the upgraded pixel detector. The 3D technology is already used in the current ATLAS detector. The main difference between the 3D and the standard planar technology is how the electrodes are built. For planar, the electrodes are incorporated in the surface of the silicon, while in 3D sensors they penetrate the bulk, forming small columns. The advantage of 3D over planar is that the distance between electrodes is decoupled from the thickness of the sensor, allowing to reduce the inter-electrode distance, meaning that the depletion voltage will be lower, the charge collection faster and the trapping probability smaller. This favours lower power dissipation, and improves radiation hardness. The main disadvantage of the 3D technology is that the fabrication process is more complicated.

### 1.1 IBL and AFP

In May 2014, a new innermost layer of the ATLAS Pixel detector called Insertable B-layer (IBL) was installed [1]. Planar and 3D silicon sensor technologies were proven to fulfil the radiation requirements of the IBL, mainly high efficiency (> 97 %) after a fluence of $5 \cdot 10^{15}$ n$_{eq}$ cm$^{-2}$ (1 MeV neutron equivalent).

The pixel sensors installed in IBL are 75% planar and 25% 3D. The 3D is produced in 230 μm thick p-type silicon substrate. They consist of an array of 336x80 pixel cells with 50x250 µm² pixel size. The columns were implanted with a double sided process and each pixel contains two n⁺ junction column (electrode distance $L_{el}$=67 µm). The sensors were designed to be bump bonded to the FEI4 readout chip [2].

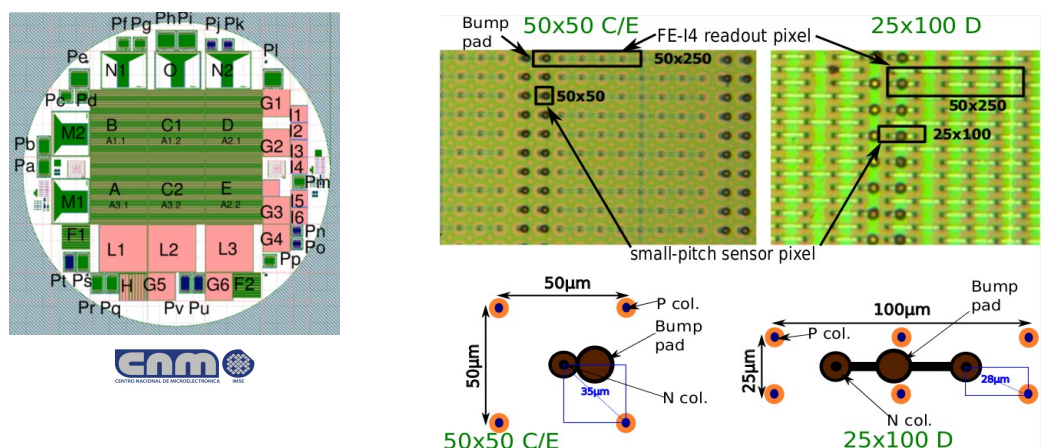

**Figure 1.** (Left) Small pitch wafer produced at CNM. The A sensor is a standard FEI4 (50x250 µm² 2E) while B (25x500 µm² 5E with guard ring), C (50x50 µm² 1E with guard ring, C1 and C2 have the same structure), D (25x100 µm² 2E) and E (50x50 µm² 1E without guard ring) have a smaller pixel size, but are adapted to be attached to the FEI4 chip. (Right) Detail of the structure for C/E and D. For them, only 20% of the sensor pixels are connected to a readout channel and the rest are grounded (top). Corresponding 3D pixel cell layout (bottom).

Another experiment that uses a tracking detector based on the 3D technology is the ATLAS Forward Proton (AFP) detector [3]. It uses IBL-like sensors with a slim edge of 180 µm that are placed at ~3 mm to the LHC beam at 210 m from the interaction point.

## 2. 3D sensors for the HL-LHC

3D sensors are being considered for the innermost layers of the upgraded Pixel detector for the HL-LHC. The IBL generation of 3D sensors needs to be upgraded to cope with the more demanding radiation conditions and higher occupancy. The high particle densities impose the need for smaller pixel sizes, which also offers increased position resolution. Furthermore, the reduced electrode distance further increases radiation hardness and the operation voltage is expected to be reduced improving the power dissipation performance.

### 2.1 Production of first 3D sensors for the HL-LHC

Two pixel sizes are being considered for the inner detector in the high luminosity upgrade, 50x50 and 25x100 µm². A first R&D production with small pixel size was fabricated at CNM[1]. Five wafers were produced with characteristics similar to the IBL generation [4]. The silicon substrate used is p-type with a resistivity of 20 kΩ cm and a thickness of 230 µm. A double sided process was used to incorporate the 200 µm deep columns with a diameter of 8 µm. The production with run number 7781 was finished in December 2015 (mask in figure 1 left).

---

[1] Centro Nacional de Microelectrónica (CNM-IMB-CSIC), 08193 Bellaterra (Barcelona), Spain.

To readout the 3D pixel sensors a more radiation hard chip than the FEI4 is being developed by the CERN RD53 collaboration [5]. Its pixel size is 50x50 μm² which is compatible with both 3D sensor small pitch geometries. It will provide thresholds of 1000 e⁻ and a lower noise level than the FEI4. The chip is still being designed and the first production is expected for 2017. Thus the first 3D sensors with new pixel sizes were designed to be tested with the FEI4 chip. To connect the new small pitch sensor pixel geometries to the FEI4 chip with a segmentation of 50x250 μm², a special layout was used. For the 50x50 μm² pixel size, one 3D n⁺ electrode was connected to one readout bump of the FEI4 while the rest were not connected to the readout, but shorted together to ground (as shown in figure 1 top centre). Two types of sensors were designed using this geometry (referred to as 50x50 1E, $L_{el}$=35 μm), one in which the full active area is surrounded by a structure of interconnected n⁺ column electrodes (3D guard ring) connected to ground (type C), and the other without guard ring (type E). For the 25x100 μm² pixel size, two geometries were considered. One has two 3D n⁺ electrodes connected together to one of the chip bumps (25x100 2E, $L_{el}$=28 μm) and the rest are connected to ground (see figure 1 top right). As the chip rows have a pitch of 50 μm, half of the sensor rows were completely connected to ground (type D). The other geometry with 25x100 μm² unit cell has five 3D n⁺ electrodes connected together to each of the FEI4 pixels (25x500 5E, $L_{el}$=52 μm). Taking two consecutive pixels of the sensor a geometry of 50x500 μm² is obtained, matching two consecutive columns on the chip, hence allows a fully active sensor (type B).

In order to perform the bump bonding cycle, an additional metal layer has to be deposited on the sensor aluminum bump pad. Such process is called under bump metallization (UBM) [6]. CNM performed an electro-less UBM process on the pixel sensors, which deposited nickel and gold on top of the aluminum pads. The electro-less process can be performed on single (diced) sensors. Two electro-less batches were produced. In the first batch, the UBM failed to deposit gold on a large fraction of bumps, resulting in large areas of disconnected pixels. For the second batch the UBM was improved and the quality of the bump bonding was greatly increased (see next section for a verification using source scans). CNM also performed an electro-plate UBM (on a full wafer), which deposited a W-Ti alloy, copper and gold. In total eight of the sensors with UBM were successfully bump bonded for R&D purposes (see table 1). Afterwards, the sensors were glued and wire bonded to a dedicated readout board. The bump bonding and assembly were made at IFAE².

It should be noted that this first production of small pitch 3D devices compatible with the FEI4 chips suffered from very low yield. Several wafers broke during or immediately after production, and the electrical characterization of the final sensors was not optimal (see next section). The production problems were traced back to damage induced in the wafer edges and to the column sidewall during the deep reaction ion etching (DRIE) process respectively. The 3D sensor production yield increased significantly after the problems were addressed [7].

**2.2 Characterization of HL-LHC 3D prototypes**

To measure the leakage current a negative voltage is applied to the back side of the sensor and the front side is connected to ground through the FEI4 chip. The measured current is representing all the connected pixels. Even with the production problem mentioned above, the sensors can be operated at least to 12 V before going into soft breakdown (see figure 2), which is enough for testing purposes, since the full depletion is expected to happen at very low voltages [8].

---

² Institut de Física d'Altes Energies (IFAE), 08193 Bellaterra (Barcelona), Spain.

| Device name | Pixel size [µm²] | Elec. Dist. [µm] | Electrodes/pixel | 3D Guard ring | UBM/Batch | Hit map |
|---|---|---|---|---|---|---|
| W4-C1 | 50x50 | 35 | 1E | yes | Electro-less 1 |  |
| W4-D | 25x100 | 28 | 2E | yes | Electro-less 1 |  |
| W8-C1 | 50x50 | 35 | 1E | yes | Electro-plate 1 |  |
| W8-C2 | 50x50 | 35 | 1E | yes | Electro-plate 1 |  |
| W8-E | 50x50 | 35 | 1E | no | Electro-plate 1 |  |
| W3-E | 50x50 | 35 | 1E | no | Electro-less 2 |  |
| W5-B | 25x500 | 52 | 5E | yes | Electro-less 2 |  |
| W3-C1 | 50x50 | 35 | 1E | yes | Electro-less 2 |  |

**Table 1.** Characteristics of successfully mounted and tested devices.

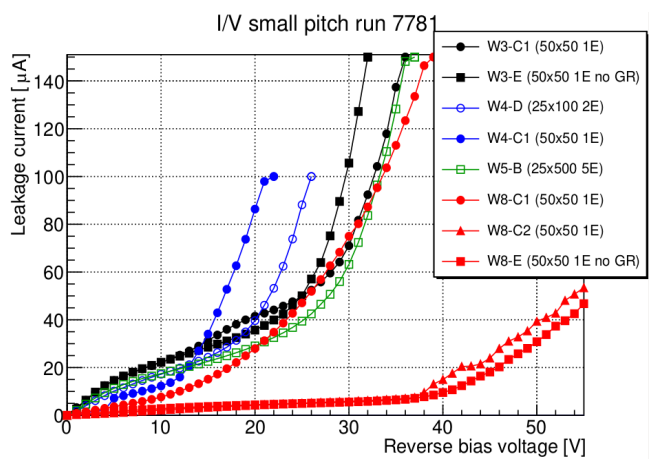

**Figure 2.** Leakage current vs. reverse bias voltage for the first 3D small pitch prototypes (run 7781) measured at room temperature. For each curve the number represents the wafer and the letter the position on the wafer corresponding to figure 1.

The charge collection properties of the different devices were studied using a $^{90}$Sr source and compared with a standard FEI4 3D sensor from the IBL production (CNM-101). The signal given by the FEI4 chip is the time over threshold from which the collected charge can be obtained (see figure 3). The source has a divergence and typically several pixels fire for each traversing particle, which are clustered together. In order to reduce charge loss to the non-sensitive sensor pixels that are not connected to the readout, events with only one firing pixel are discarded. The most probable value (MPV) of the cluster charge is obtained from a Landau-Gauss fit (see figure 3 bottom left). A calibration factor using an $^{241}$Am gamma source was applied. The expected MPV charge is 16.8 ke$^-$. The measurements show that for most of the devices the collected charge is around the expected value and similar to the standard FEI4 refer-ence device even at 0 V (within calibration uncertainties of ~20%). For two sensors, W4-D and W8-C1, the charge collected is lower than expected. As mentioned before, to adapt the D geometry to the FEI4 chip each 25x100 μm² pixel only has neighbours in one direction, hence more charge is lost to the insensitive regions than for the other geometries. In the case of W8-C1 the loss of charge can be explained with the map of connected pixels. A large area of the sensor is connected, but there are many individual disconnected pixels in between where the charge is not read-out and hence charge loss can happen.

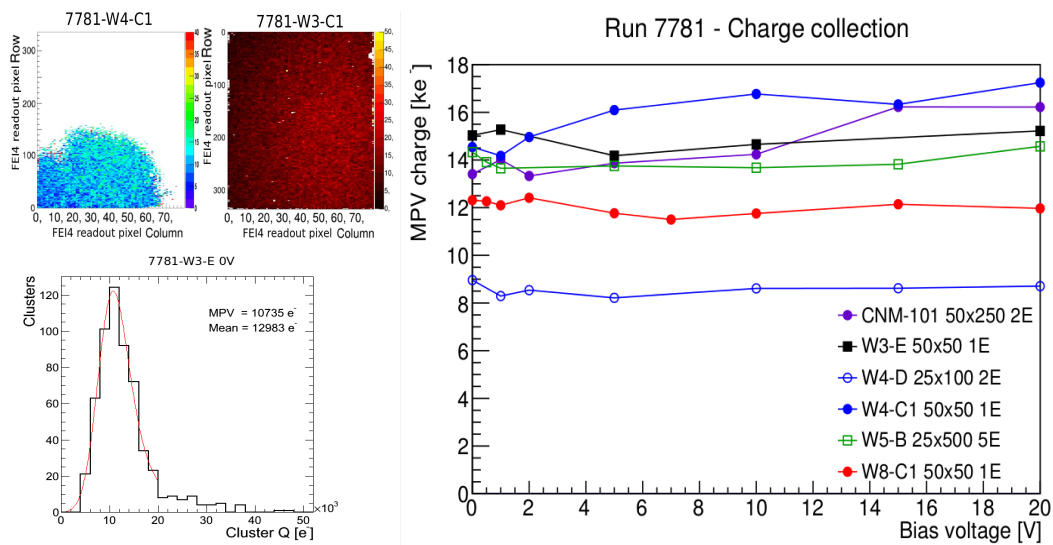

**Figure 3.** (Top left) $^{90}$Sr source scan hit map for two of the sensors with electro-less UBM, one of the first UBM batch (left) and one of the second batch (right). (Bottom left) Clustered charge distribution fitted with a Landau-Gauss. (Right) Collected charge of the small pitch sensors compared to an IBL FEI4 3D sensor (CNM-101).

## 3. 3D sensors at HL-LHC fluences

As already mentioned, the HL-LHC pixel tracker will require a detector able to cope with high radiation levels. The maximum expected particle fluence is 1 to $2 \cdot 10^{16}$ $n_{eq}$ cm$^{-2}$ in the innermost layer [9]. The leakage current is an important parameter to study in irradiated devices, since it evolves with increasing fluence and can limit the detector operability either by introducing too much noise or by reaching the limit of the voltage supply. Also thermal runaway can occur since the current has a strong dependence on temperature, and the dissipated power increases the temperature, possibly entering in a feedback cycle. The devices need cooling in order to dissipate the heat and to avoid the effect of this feedback. The leakage current was measured for two different classes of devices irradiated with neutrons: 3D pixel sensors from the IBL generation compatible with the FEI3 chip, and strip sensors from the new small pitch production. The measurements were performed in a climate chamber at -25 ºC, hence the results shown here are upper limits since in the experiment the sensors are cooled more efficiently.

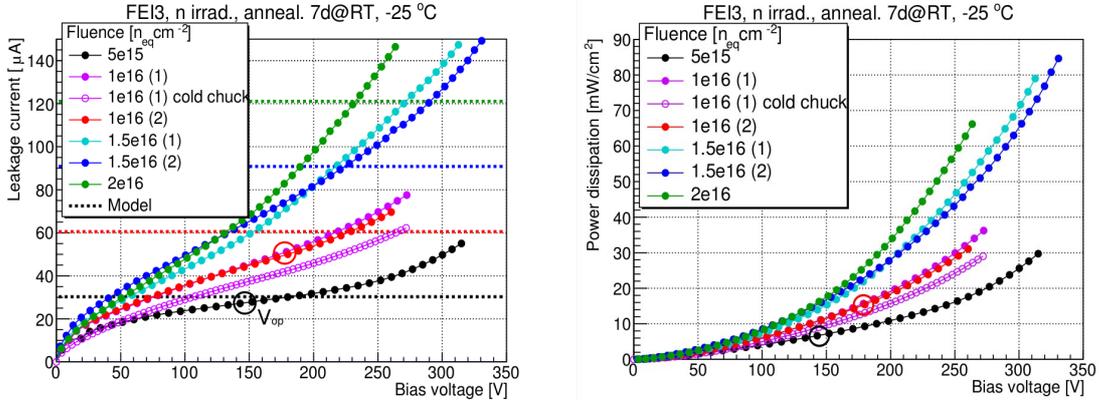

**Figure 4.** Leakage current (left) and power dissipation (right) of FEI3 sensors from the IBL production irradiated up to HL-LHC fluences, measured in a climate chamber at -25 ºC. Two different sensors were measured at $10^{16}$ and $1.5 \cdot 10^{16}$ $n_{eq}$ cm$^{-2}$ (1 and 2). The discontinuous lines correspond to the expected current assuming it is described by radiation-induced bulk damage. For a fluence of $10^{16}$ $n_{eq}$ cm$^{-2}$, sensor 1 was also measured in contact with a cold chuck (open points). The operation voltage is shown in a circle for the smaller fluences.

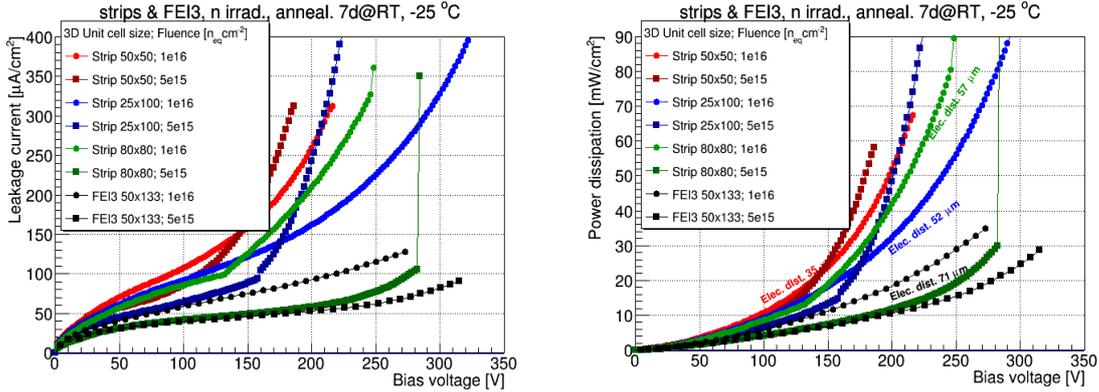

**Figure 5.** Leakage current per area (left) and power dissipation (right) of small pitch irradiated strips, compared to strips and FEI3 sensors from the IBL production for fluences of $5 \cdot 10^{15}$ and $10^{16}$ $n_{eq}$ cm$^{-2}$. The electrode distance is specified.

### 3.1 IBL generation sensors at HL-LHC fluences

The FEI3 chip is an older version of the FEI4 used in the ATLAS Pixel detector. It consists of an array of 160x18 pixels with a pixel size of 50x400 μm². The 3D FEI3 sensors have 3 electrodes connected per pixel (50x400 3E, $L_{el}$=71 μm). The sensors were irradiated with neutrons at JSI Ljubljana to fluences of $5 \cdot 10^{15}$, $1 \cdot 10^{16}$, $1.5 \cdot 10^{16}$ and $2 \cdot 10^{16}$ $n_{eq}$ cm$^{-2}$. The FEI3 was used because the FEI4 chip has tantalum which becomes highly radioactive after neutron irradiation. They were mounted on a read-out board and annealed during one week at room temperature. The leakage current was measured applying a voltage to the back side of the sensor and connecting the front side through the FEI3 chip to ground. Previous measurements at lower fluences show a plateau starting at the voltage where the sensor is fully depleted. However, for the high fluences measured here, no real plateau is visible; only for $5 \cdot 10^{15}$ $n_{eq}$ cm$^{-2}$ there is some region of reduced slope. The expected dependence of the leakage current with fluence is linear at low fluences if dominated by the radiation effects in the bulk [10]. In theory, the current on the plateau should match with the model after depletion. Following the linear dependence up to the measured fluences the leakage current was calculated and compared with the current of the FEI3 (see figure 4). For the fluences of $5 \cdot 10^{15}$ and $10^{16}$ $n_{eq}$ cm$^{-2}$ the leakage current is in the range of the expected values at the operation voltages determined from hit efficiency in beam tests (145 and 180 V respectively [11]). The power dissipation at these voltages are 7 and 15 mW/cm² respectively. For the higher fluences the operation voltage still needs to be determined.

In addition, one of the FEI3 sensors irradiated to a fluence of $1 \cdot 10^{16}$ $n_{eq}$ cm$^{-2}$ was measured without board directly on a cold chuck. The chip in contact with the chuck dissipates the heat better than in the climate chamber, hence the temperature of the chip is reduced more efficiently and the measured leakage current is lower. The comparison between the measurements show a difference in leakage current of 20%.

### 3.2 3D small pitch strips at HL-LHC fluences

The strip sensors measured from the small pitch production have 3D unit cells of 50x50 and 25x100 μm² ($L_{el}$=35 and 57 μm respectively). Strips were studied because their production and testing is easier than for the pixel sensors. They were measured together with strips of 80x80 μm² ($L_{el}$=57 μm) 3D unit cell from the IBL production, and compared to the FEI3 from section 3.1. The aim is to compare the leakage current and power dissipation of the new structures with the measurements of the IBL generation devices.

The strips were irradiated with neutrons at JSI Ljubljana to fluences of $5 \cdot 10^{15}$ and $10^{16}$ $n_{eq}$ cm$^{-2}$ and annealed for one week at room temperature. They were glued and wire bonded to a read-out board where the leakage current of all the strips can be measured together. It was scaled by area to compare devices of different sizes.

The results show a higher leakage current for smaller electrode distance at the same fluence (see figure 5). It could be an effect of charge multiplication in higher electric fields at smaller electrode distances, or an artifact of this first small-pitch 3D run with non-ideal leakage currents already before irradiation. Further studies are on-going. Nevertheless, the operation voltages for small pitch sensors are expected to be lower for smaller 3D cell sizes, as described above, which will have a compensating effect. The determination of the operation voltages for the small pitch sensors will be done in beam tests. The small pitch strips can be operated up to 150 V with reasonably low power dissipation (< 25 mW/cm² at $10^{16}$ $n_{eq}$ cm$^{-2}$).

### 4. Conclusions

3D silicon sensors are a mature technology used in the ATLAS IBL and AFP experiments. However, for the HL-LHC high radiation hardness and small pixel sizes are required. The first prototypes of small pitch sensors have been successfully produced at CNM and tested at IFAE. They can be operated at least up to 12 V. The collected charge is similar to the standard FEI4 3D sensors and close to the expected value of 16.8 ke$^-$, even at very low operation voltages of 0 V. The leakage current of the IBL-type 3D sensors was tested up to a fluence of $2 \cdot 10^{16}$ $n_{eq}$ cm$^{-2}$. The current roughly matches the linear dependence with fluence at the operation voltage. The leakage current of the first small pitch strips irradiated is higher than for IBL-like devices, but the expected lower operation voltage may compensate the effect on power dissipation. Beam tests with the small pitch 3D pixel sensors before and after irradiation were carried out and the analysis is on-going. A run with the same mask as the small pitch run 7781 presented here, but with improved process is under production. A much higher yield and better breakdown voltages are expected. In addition, a single sided run with thinner wafers between 70 and 150 μm is on-going. A small pixel run adapted to the RD53 chip is also being produced.

**Acknowledgments**

The authors wish to thank V. Cindro and I. Mandic (JSI Ljubljana) for excellent support for the irradiations. This work was partly performed in the framework of the CERN RD50 collaboration. This work was partially funded by: the MINECO, Spanish Government, under grants FPA2013-48308-C2-1-P, FPA2015-69260-C3-2-R, FPA2015-69260-C3-3-R and SEV-2012-0234 (Severo Ochoa excellence programme); and the European Union's Horizon 2020 Research and Innovation programme under Grant Agreement no. 654168.

**References**

[1] ATLAS IBL Collaboration, *Prototype ATLAS IBL Modules using the FE-I4A Front-End Readout Chip*, JINST 7 (2012) P11010.

[2] M. Garcia-Sciveres et al., *The FE-I4 pixel readout integrated circuit,* NIM A 636 (2011) S155.

[3] J. Lange et al., *3D silicon pixel detectors for the ATLAS Forward Physics experiment,* JINST 10 (2015) C03031, arXiv:1501.02076 [physics.ins-det].

[4] G. Pellegrini et al, *First double-sided 3-D detectors fabricated at CNM-IMB*, NIM A Volume 592 Issues 1-2, 11 July 2008, Pages 38-43 DOI: 10.1016/j.nima.2008.03.119

[5] RD53 Collaboration, *RD Collaboration Proposal: Development of pixel readout integrated circuits for extreme rate and radiation,* CERN-LHCC-2013-008 (2013).

[6] M. Bigas, E. Cabruja, M. Lozano, *Bonding techniques for hybrid active pixel sensors (HAPS)*, NIM A Volume 574, Issue 2, 1 May 2007, Pages 392-400, ISSN 0168-9002.

[7] S. Grinstein et al., *Experience with the AFP 3D Silicon Pixel Tracker*, in Proceedings of the 8$^{th}$ International Workshop on Semiconductor Pixel Detectors for Particles and Imaging, to be published in JINST. September, 2016.

[8] M. Baselga, *Development of pixel detectors for the IBL and HL-LHC ATLAS experiment upgrade* PhD Thesis.

[9] A. Clark, M. Elsing, P. Wells, *Performance Specifications of the Tracker Phase II Upgrade,* ATL-UPGRADE-PUB-2012-003, ATL-COM-UPGRADE-2012-022

[10] M. Moll, *Radiation Damage in Silicon Particle Detectors* PhD Thesis, DESY-THESIS-1999-040.

[11] J. Lange et al., *3D silicon pixel detectors for the High-Luminosity LHC,* submitted to JINST, October 2016, arXiv:1610:07480 [physics.ins-det].